\documentclass[twocolumn]{revtex4}

\usepackage{graphicx}
\newcommand{\RT}[1]{``#1,''}

\newcommand{\CiteQM}{QM_OurPRL2013, QM_OurPRL2018, QM1, QM2, QM3, QM4, QM5}

\newcommand{\CiteQPG}{QPG1, QPG2, QPG3, QPG4, QPG5, QPG6, QPG7, QPG8}

\newcommand{\CiteQFC}{QFC1, QFC_OurPRR2021, QFC2, QFC3}

\newcommand{\CiteQuantumCommunication}{QC1, QC2, QC3, QC4, QC5}

\newcommand{\CiteSQ}{SQ1, SQ2, SQ3}

\newcommand{\CiteQN}{QN1, QN2, QN3}

\begin{document}
\title{Temporally-ultralong biphotons with a linewidth of 50 kHz
}

\author{Yu-Sheng Wang,$^1$ Kai-Bo Li,$^1$ Chao-Feng Chang,$^1$ Tan-Wen Lin,$^1$ Jian-Qing Li,$^1$ Shih-Si Hsiao,$^1$ Jia-Mou Chen,$^1$ Yi-Hua Lai,$^1$ Ying-Cheng Chen,$^{2,4}$ Yong-Fan Chen,$^{3,4}$ Chih-Sung Chuu,$^{1,4}$ 
Ite A. Yu$^{1,4,\ast}$}

\address{$^1$Department of Physics, National Tsing Hua University, Hsinchu 30013, Taiwan \\
$^2$Institute of Atomic and Molecular Sciences, Academia Sinica, Taipei 10617, Taiwan\\
$^3$Department of Physics, National Cheng Kung University, Tainan 70101, Taiwan \\
$^4$Center for Quantum Technology, Hsinchu 30013, Taiwan \\
~~~~$^\ast$Corresponding~author:~yu@phys.nthu.edu.tw~~~~
}

\begin{abstract}
We report the generation of biphotons, with a temporal full width at the half maximum (FWHM) of 13.4$\pm$0.3 $\mu$s and a spectral FWHM of 50$\pm$1 kHz, via the process of spontaneous four-wave mixing. The temporal width is the longest, and the spectral linewidth is the narrowest up to date. This is also the first biphoton result that obtains a linewidth below 100 kHz, reaching a new milestone. The very long biphoton wave packet has a signal-to-background ratio of 3.4, which violates the Cauchy-Schwarz inequality for classical light by 4.8 folds. Furthermore, we demonstrated a highly-tunable-linewidth biphoton source and showed that while the biphoton source's temporal and spectral width were controllably varied by about 24 folds, its generation rate only changed by less than 15\%. A spectral brightness or generation rate per pump power per linewidth of 1.2$\times$10$^6$ pairs/(s$\cdot$mW$\cdot$MHz) was achieved at the temporal width of 13.4 $\mu$s. The above results were made possible by the low decoherence rate and high optical depth of the experimental system, as well as the nearly phase-mismatch-free scheme employed in the experiment. This work has demonstrated a high-efficiency ultranarrow-linewidth biphoton source, and has made a substantial advancement in the quantum technology utilizing heralded single photons.
\end{abstract}

\maketitle

\newcommand{\Table}{
	\begin{table*}[t]
	\caption{Biphoton sources with spectral profiles of the full width at the half maximum (FWHM) below 1~MHz.}
	{\centering
		\begin{tabular}{c c c c c c}
			\hline\hline
			\parbox{13mm}{\centering Process} & 
			\parbox{28mm}{\centering Medium} &
			\parbox{13mm}{\centering Type$^{(a)}$} &
			\parbox{18mm}{\centering \vspace*{1mm}Temporal\\FWHM\\($\mu$s)\vspace*{1mm}} & 
			\parbox{18mm}{\centering \vspace*{1mm}Spectral\\FWHM\\(kHz)\vspace*{1mm}} &
			\parbox{13mm}{\centering Reference} \\ 
 			\hline 
 			SPDC & nonlinear crystal & MM & 0.33$^{(b)}$ & 670$^{(c)}$
				& \cite{SPDC.Cavity10} \\ 
 			SPDC & nonlinear crystal & MM & 0.83$^{(b)}$ & 265$^{(c)}$ 
				& \cite{LongBiphotonSPDC.Cavity} \\ 
  			SFWM & cold atom cloud & SM & 1.7 & 430 
				& \cite{SFWM.dLambda.ColdAtoms7.Du} \\			
			SFWM & cold atom cloud & SM & 2.1 & 380 
				& \cite{SFWM.dLambda.ColdAtoms9.Zhang}\\  
			SFWM & cold atom cloud & SM & 2.9 & 250
				& \cite{SFWM.dLambda.ColdAtoms10.Du} \\
			SFWM & cold atom cloud & SM & 13.4 & 50
				& this work \\
 			SFWM & hot atomic vapor & SM & 0.66 & 320
				& \cite{OurOpEx2021, OurPRR2022} \\ 
			\hline\hline
		\end{tabular}
	\\ \vspace*{-0.5\baselineskip}
	\parbox{125mm}{
	\begin{flushleft}
		$^{(a)}$MM denotes multimode, and SM denotes single mode.\\
		$^{(b)}$The FWHM of the envelope formed by all peaks in a temporally comb-like structure. \\
		$^{(c)}$The spectral FWHM of the envelope. \\
	\end{flushleft}
	}
	}
	\label{TableOne}
	\end{table*}
}

\newcommand{\FigOne}{
	\begin{figure*}[t]
	\center{\includegraphics[width=135mm]{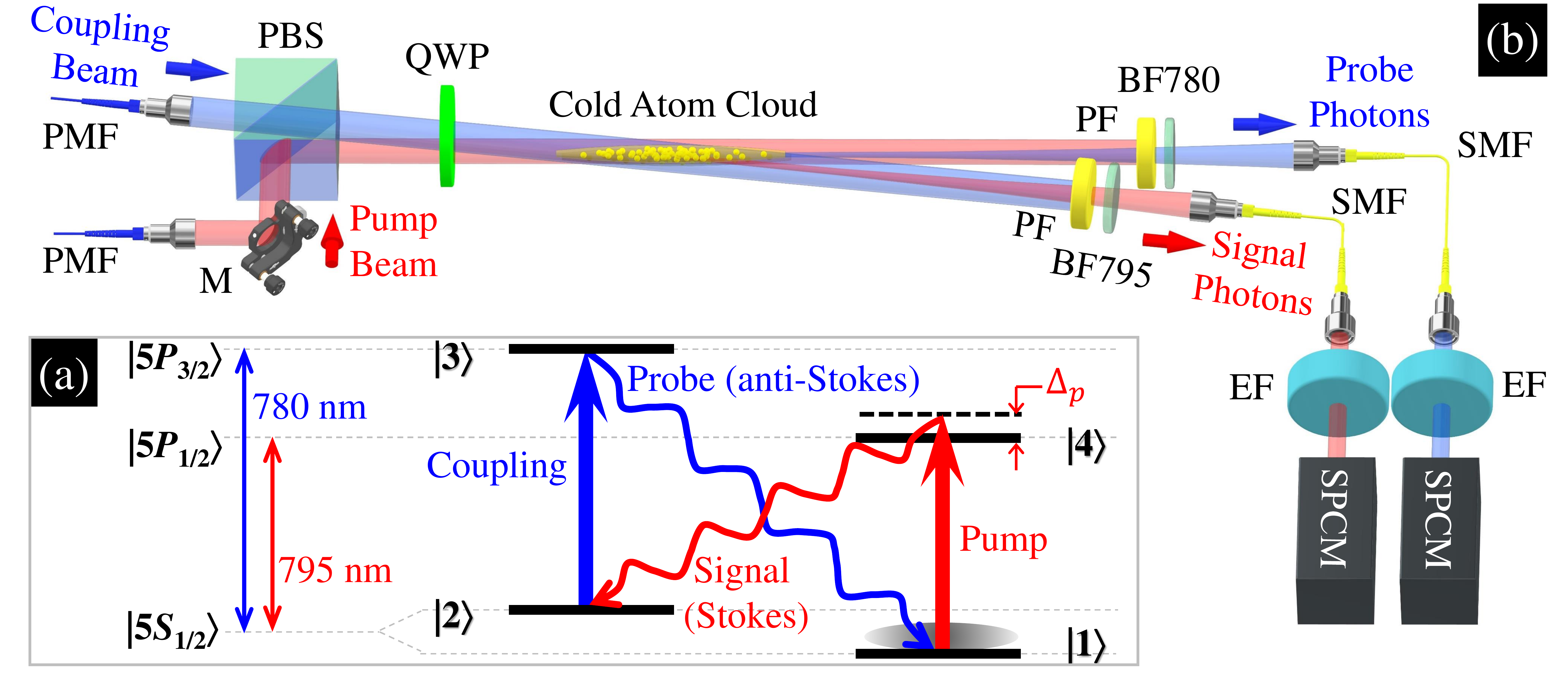}}
	\caption{
(a) 
Relevant energy levels of $^{87}$Rb atoms and the transition diagram in the experiment. Ground states $|1\rangle$ and $|2\rangle$ and excited states $|3\rangle$ and $|4\rangle$ are the Zeeman states of $|5S_{1/2},F=1,m=1\rangle$, $|5S_{1/2},F=2,m=1\rangle$, $|5P_{3/2},F=2,m=2\rangle$, and $|5P_{1/2},F=2,m=0\rangle$, respectively. The pump field was blue-detuned with a detuning $\Delta_p$ of 200~MHz. The signal (probe) photons are also called the Stokes (anti-Stokes) photons. The frequency difference between $|1\rangle$ and $|2\rangle$ is 6.8~GHz, corresponding to a wavelength difference of merely 0.014~nm.
(b)
Schema of the experimental setup. The angle separation between the pump and coupling propagation directions was approximately 1$^{\circ}$. The 795-nm signal photons and the 780-nm coupling field (the 780-nm probe photons and the 795-nm pump field) propagated in the same direction. PMF: polarization-maintained optical fiber, M: mirror, PBS: polarizing beam splitter, QWP: quarter-wave plate, BF780 (BF795): 780 nm (795 nm) bandpass filter, PF: polarization filter, SMF: single-mode optical fiber, EF: etalon filter, and SPCM: single-photon counting module.
	}
	\label{fig:transitions}
	\label{fig:setup}
	\end{figure*}
}
\newcommand{\FigTwo}{
	\begin{figure}[t]
	\center{\includegraphics[width=75mm]{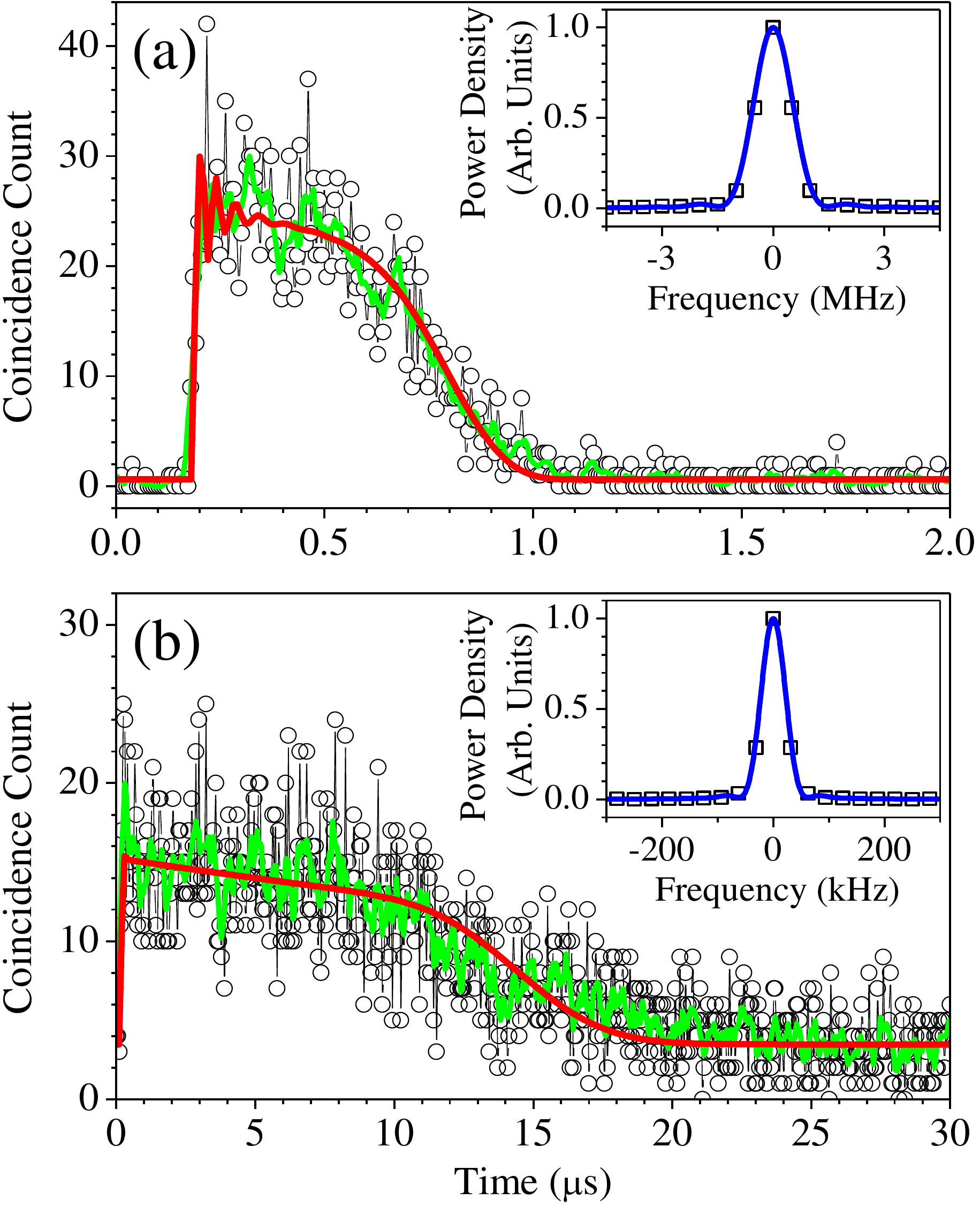}}
	\caption{
Representative biphoton wave packets. Circles connected with black lines are the data of two-photon coincidence counts. Red lines are the theoretical predictions of $G^{(2)}(\tau)$ in Eq.~(\ref{eq:biphoton}). Green lines are the results of four-point moving average on the data points. In each inset, squares represent the discrete Fourier transform of the data and blue line is the Fourier transform of the red line. (a) The coupling power was 1.5~mW, and the pump power and detuning were 56~$\mu$W and +200~MHz. Red line corresponds to $(\alpha, \Omega_c, \gamma)$ = $(115, 2.1\Gamma, 4.0{\times}10^{-3}\Gamma)$ and $(\Omega_p, \Delta_p)$ = $(0.32\Gamma, 33.3\Gamma)$. The biphoton wave packet had a temporal full width at the half maximum (FWHM) of 0.57~$\mu$s, a spectral FWHM of 1.20~MHz, and a signal-to-background ratio (SBR) of 40. (b) The coupling power was reduced to 63~$\mu$W, while the pump power and detuning were the same. Red line corresponds to $(\alpha, \Omega_c,  \gamma)$ = $(110, 0.42\Gamma, 3{\times}10^{-4}\Gamma)$ and $(\Omega_p, \Delta_p)$ being the same. The biphoton wave packet had a temporal FWHM of 13.4~$\mu$s, a spectral FWHM of 50~kHz, and a SBR of 3.4.
	}
	\label{fig:biphoton}
	\end{figure}
}
\newcommand{\FigThree}{
	\begin{figure}[t]
	\center{\includegraphics[width=80mm]{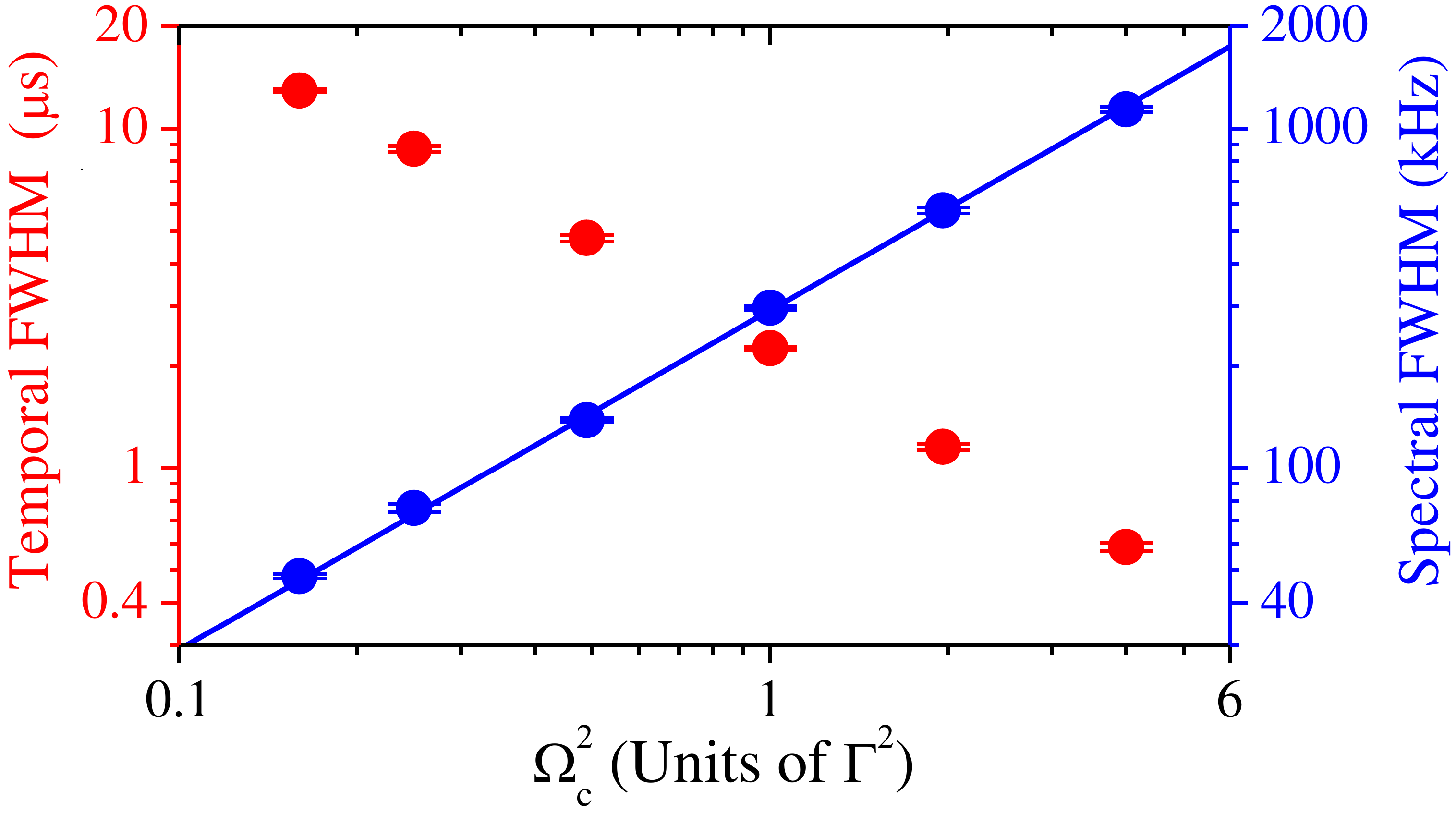}}
	\caption{Temporal FWHM (red circles) and spectral FWHM (blue circles) of the biphoton wave packet as functions of $\Omega_c^2$. In the measurement, $\alpha$ = 110$\pm$5. Blue line is the best fit of a linear function with zero interception, which determines $\Delta \omega /2\pi$ (the spectral FWHM) = 280~kHz$\times$$(\Omega_c/\Gamma)^2$. In theory, i.e., Eq.~(\ref{eq:linewidth}), it is expected that $\Delta \omega /2\pi =$ 300~kHz$\times$$(\Omega_c/\Gamma)^2$ at $\alpha = 110$.
	}
	\label{fig:width}
	\end{figure}
}
\newcommand{\FigFour}{
	\begin{figure}[t]
	\center{\includegraphics[width=85mm]{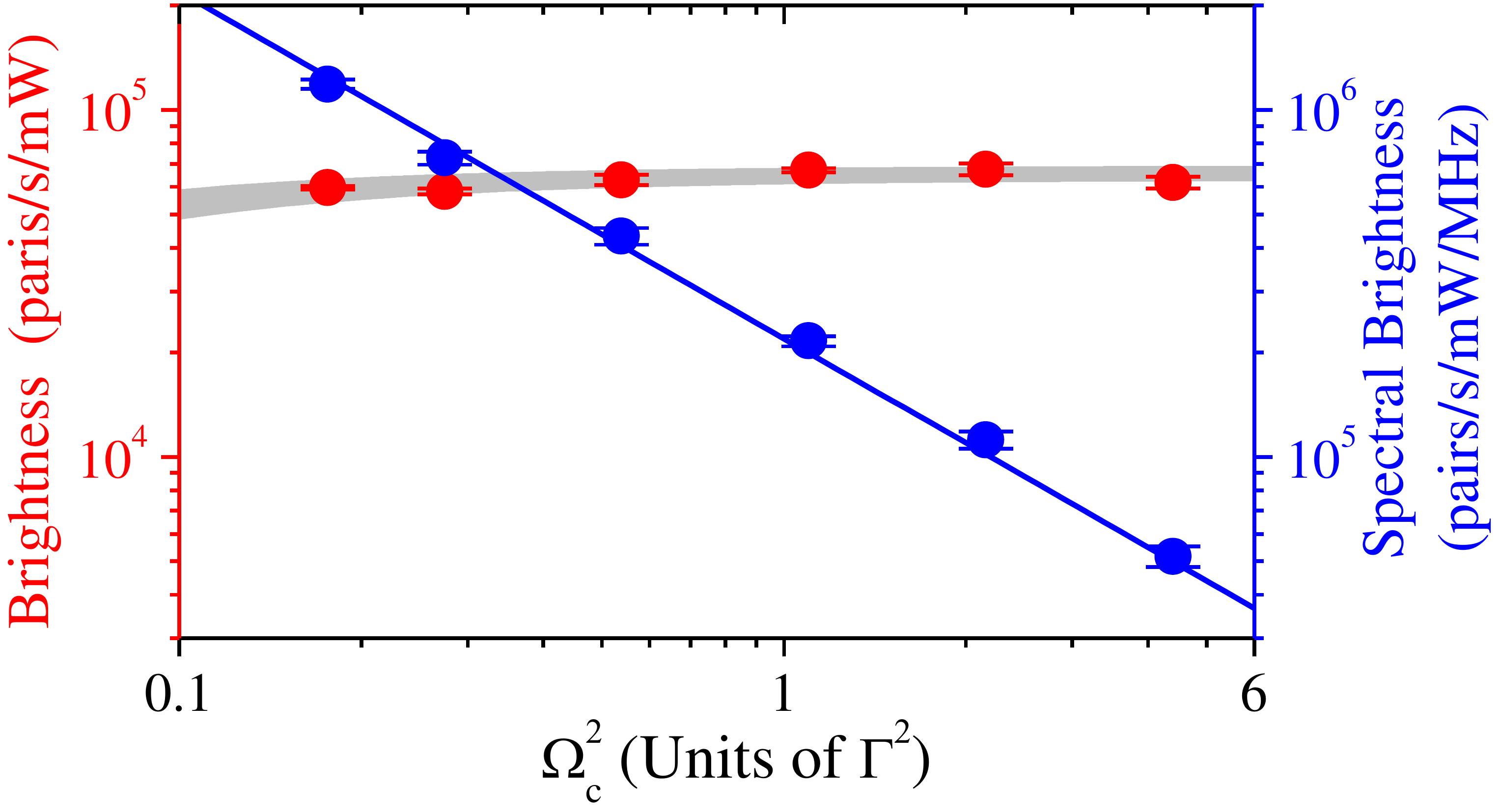}}
	\caption{Brightness, i.e., the generation rate per pump power, (red circles) and spectral brightness, i.e., the brightness per linewidth, (blue circles) as functions of $\Omega_c^2$. The pump power was kept at 56~$\mu$W, and the generation rates of different coupling Rabi frequencies varied in the range of 3300$\sim$3800 pairs/s. During the entire measurement, the value of $\gamma$ was (3$\sim$40)$\times$$10^{-4}$$\Gamma$ corresponding to $\Omega_c =$ (0.4$\sim$2.0)$\Gamma$, and that of $\alpha$ was bewteen 105 and 115. Considering the uncertainty or fluctuation in $\gamma$ of $\pm$20\% and that in $\alpha$ of $\pm$5, gray area represents the predictions based on Eq.~(\ref{eq:GR}). Blue line is the best fit of the function $y = b/x$, where $b$ is the fitting parameter.
	}
	\label{fig:brightness}
	\end{figure}
}
\section{Introduction}

Sources of biphotons or pairs of time-correlated single photons are widely utilized in quantum information research and applications. While one of the paired photons is detected to trigger, or start, a quantum operation, the other will be employed in the operation as a heralded single photon or, if carrying a wave function or quantum state, a heralded qubit. Biphotons are produced mainly via two kinds of schemes: the spontaneous parametric down conversion (SPDC) and the spontaneous four-wave mixing (SFWM). A SPDC biphoton source is commonly made of a nonlinear crystal placed inside an optical cavity 
\cite{
SPDC.Cavity2, SPDC.Cavity5, SPDC.Cavity6, SPDC.Cavity9, SPDC.Cavity10, SPDC.Cavity11, SPDC.Cavity12, SPDC.Cavity13, SPDC.Cavity.Review, LongBiphotonSPDC.Cavity
}. 
A SFWM biphoton source generally consists of a cold atom cloud 
\cite{
SFWM.dLambda.ColdAtoms7.Du, SFWM.dLambda.ColdAtoms9.Zhang, SFWM.dLambda.ColdAtoms10.Du, SFWM.dLambda.ColdAtoms11, SFWM.dLambda.ColdAtoms12, SFWM.dLambda.ColdAtoms13, 
SFWM.ladder.ColdAtoms2
}, 
a hot atomic vapor 
\cite{
SFWM.dLambda.HotAtom1, SFWM.dLambda.HotAtom2, SFWM.dLambda.HotAtom5, SFWM.dLambda.HotAtom6, OurOpEx2021, SFWM.dLambda.HotAtom7, OurPRR2022,
SFWM.ladder.HotAtom5, SFWM.ladder.HotAtom7, SFWM.ladder.HotAtom8, SFWM.ladder.HotAtom9
}, 
or an integrated photonics device with a built-in micro-resonator or waveguide 
\cite{
IPD7, IPD8, IPD9, IPD10, IPD11, IPD12, IPD.Review2, IPD.Review3, LongBiphotonIPD
}. 
Both the SPDC biphoton source and the micro-resonator/waveguide SFWM biphoton source can achieve a high generation rate 
\cite{
SPDC.Cavity2, SPDC.Cavity5, SPDC.Cavity6, SPDC.Cavity9, SPDC.Cavity10, SPDC.Cavity11, SPDC.Cavity12, SPDC.Cavity13, SPDC.Cavity.Review, LongBiphotonSPDC.Cavity, 
IPD7, IPD8, IPD9, IPD10, IPD11, IPD12, IPD.Review2, IPD.Review3, LongBiphotonIPD
}. 
On the other hand, the atomic SFWM biphoton source, which utilizes the effect of electromagnetically induced transparency (EIT), is able to accomplish a long correlation time between the pair of single photons, i.e., a long temporal width as well as a narrow spectral linewidth of the biphoton wave packet 
\cite{
SFWM.dLambda.ColdAtoms7.Du,  SFWM.dLambda.ColdAtoms9.Zhang, SFWM.dLambda.ColdAtoms10.Du, SFWM.dLambda.ColdAtoms11, SFWM.dLambda.ColdAtoms12, SFWM.dLambda.ColdAtoms13, 
SFWM.dLambda.HotAtom1, SFWM.dLambda.HotAtom2, SFWM.dLambda.HotAtom5, SFWM.dLambda.HotAtom6, OurOpEx2021, SFWM.dLambda.HotAtom7, OurPRR2022
}.

A quantum operation employing biphotons of a narrower linewidth or, equivalently, a longer temporal width, can achieve a better efficiency. For example, a narrower linewidth of input photons can result in a higher storage efficiency for quantum memory \cite{\CiteQM}, a greater success rate for quantum phase gates \cite{\CiteQPG}, and a larger conversion efficiency for quantum frequency converters \cite{\CiteQFC}, which utilize resonant or quasi-resonant transition schemes. As another example, while the entanglement swapping is a key process in the quantum repeater protocol for long-distance quantum communication \cite{\CiteQuantumCommunication}, the deterministic entanglement swapping with the fidelity of 79\% has been demonstrated with ion qubits \cite{DES2008}, which operated at a narrow-linewidth transition. In addition, the EIT effect has recently been observed in superconducting qubits or artificial atoms driven by narrow-linewidth microwaves  \cite{\CiteSQ}. One can foresee that narrow-linewidth photonic qubits, which are converted to and from microwave coherences based on the EIT effect \cite{\CiteQM}, will be employed in a quantum network that links superconducting-qubit-based quantum computers \cite{\CiteQN}.

\Table

Five groups have reported the sources of biphotons with sub-MHz linewidths. Utilizing nonlinear crystals in the cavity-assisted SPDC process, Rambach et.\ {\it al}.\ showed that their biphotons had a linewidth of 430~kHz \cite{SPDC.Cavity10}, and Liu et.\ {\it al}.\ demonstrated a biphoton source with a linewidth of 265~kHz \cite{LongBiphotonSPDC.Cavity}. Both sources produced multi-mode biphotons, and the frequency modes of the biphotons spanned several hundreds of MHz. Although SPDC biphoton sources can generate single-mode biphotons, currently all of these single-mode biphotons have linewidths larger than 1~MHz, e.g., Refs.~\cite{SPDC.Cavity10, SPDC.Cavity11, SPDC.Cavity12}. Using laser-cooled atoms in the EIT-based SFWM process, Han et.\ {\it al}.\ achieved a biphoton source with a linewidth of 380~kHz \cite{SFWM.dLambda.ColdAtoms9.Zhang}, and Zhao et.\ {\it al}.\ reported that their biphotons had a linewidth of 430~kHz \cite{SFWM.dLambda.ColdAtoms7.Du} or 250~kHz \cite{SFWM.dLambda.ColdAtoms10.Du}. Employing an atomic vapor heated to 38$^{\circ}$C in the EIT-based SFWM process, Hsu et.\ {\it al}.\ generated biphotons with a linewidth of 320~kHz \cite{OurOpEx2021, OurPRR2022}. Among all the biphoton sources of integrated photonics devices, the narrowest linewidth was 92~MHz \cite{LongBiphotonIPD}. There has been no report on the biphoton source with a linewidth below 100~kHz until now. Please see Table \ref{TableOne} for the list of all biphoton sources with linewidths below 1~MHz.

Here, we report a cold-atom SFWM source of biphotons with a tunable temporal width. While maintaining non-classicality, the biphotons had a temporal width as long as 13.4~$\mu$s, or a spectral linewidth as narrow as 50~kHz. A large optical depth (OD) as well as a negligible decoherence rate in the experimental system enabled the propagation delay time of the EIT effect to dominate the temporal profile of the biphoton wave packet. Furthermore, we were able to tune the temporal or, equivalently, spectral width by about 24 folds, and at the same time the change in the generation rate of the biphoton source was less than 15\%. A generation rate per pump power per linewidth of 1.2$\times$10$^6$ pairs/(s$\cdot$mW$\cdot$MHz) was achieved at the temporal width of 13.4~$\mu$s, due to the nearly phase-mismatch-free scheme employed in our experiment. In quantum operations, this biphoton source can meet any harsh requirements of linewidth.

\section{Experimental Setup}

\FigOne

We carried out the exepriment in cold $^{87}$Rb atoms, which were produced by a magneto-optical trap (MOT). Details of the cold atoms and the MOT can be found in Ref.~\cite{QFC_OurPRR2021}. Before each measurement of biphotons, we momentarily performed the dark-MOT and optically pumped all population to the single Zeeman state of $|5S_{1/2},F=1,m=1\rangle$, i.e., the state $|1\rangle$ in Fig.~\ref{fig:transitions}(a). The processes of the dark-MOT and the optical pumping are very similar to those described in our previous works of Refs.~\cite{OurPRL2012, QM_OurPRL2013}. At the end of the processes, the optical depth (OD) of the system was 110$\pm$5 throughout this work. 

The transition diagram of the spontaneous four-wave mixing (SFWM) process is shown in Fig.~\ref{fig:transitions}(a). Since all population was optically pumped to $|1\rangle$, only the four Zeeman states specified in the caption were relevant to the experiment. The pump field had the $\sigma-$ polarization and drove the transition from $|1\rangle$ to $|4\rangle$ with a blue detuning of 200 MHz. The coupling field had the $\sigma+$ polarization and drove the transition from $|2\rangle$ to $|3\rangle$ resonantly. In the SFWM process, a pair of signal and probe photons were emitted by the transitions of $|4\rangle \rightarrow |2\rangle$ and $|3\rangle \rightarrow |1\rangle$, and had the polarizations of $\sigma_-$ and $\sigma_+$, respectively. The frequencies of the pump field, signal photon, coupling field, and probe photon maintain the four-photon resonance. Since the energy level of $|1\rangle$ is lower than that of $|2\rangle$, the signal and probe photons are also called the Stokes and anti-Stokes photons. The pump field and Stokes photons formed the Raman transition scheme, and their wavelengths were about 795~nm. The coupling field and anti-Stokes photons formed the electromagnetically-induced transparency (EIT) transition scheme, and their wavelengths were about 780~nm.


The sketch of the experimental setup is depicted in Fig.~\ref{fig:setup}(b). We merged the pump and coupling fields with a polarizing beam splitter (PBS). The two fields propagated in the directions with a small angle separation of about $1^{\circ}$. After the PBS, a quarter-wave plate made the pump and coupling fields become $\sigma-$ and $\sigma+$ polarized. Their beam profiles completely overlapped with the cigar-shaped atom cloud \cite{QFC_OurPRR2021}. The pump and coupling fields had the same $e^{-2}$ full widths of 2.0 mm. In this work, we set the pump power to 56~$\mu$W, and varied the coupling power from 60~$\mu$W to 1.5~mW. The Rabi frequency of the pump field was estimated according to the Gaussian beam width and the transition dipole matrix element of $|5S_{1/2},F=1,m=1\rangle$ $\rightarrow$ $|5P_{1/2},F=1,m=0\rangle$. The peak intensity of the 56~$\mu$W pump beam corresponds to the Rabi frequency of 0.32$\Gamma$.  The determination of the coupling Rabi frequency is illustrated in the next paragraph.

We experimentally determined the quoted parameters of coupling Rabi frequency ($\Omega_c$), optical depth or OD ($\alpha$), and decoherence rate ($\gamma$) \cite{QFC_OurPRR2021, OurPRA2019}. First, the EIT spectrum of a weak probe field was measured, and the separation distance between two transmission minima, i.e. the Autler-Townes splitting, determined $\Omega_c$. In the spectrum measurement, the OD was intentionally reduced such that the two minima can be clearly observed. The Rabi frequency of the 60~$\mu$W (or 1.5~mW) coupling beam corresponded to $\Omega_c$ of 0.4$\Gamma$ (or 2.0$\Gamma$). Knowing the value of $\Omega_c$, we next measured the delay time of a short probe pulse to determine the $\alpha$ used in the experiment. Finally, we used the values of $\Omega_c$ and $\alpha$ as well as measured the peak transmission of a long probe pulse to determine $\gamma$. This input probe pulse was long enough that its output is not affected by the EIT bandwidth. Once $\Omega_c$, $\alpha$, and $\gamma$ were determined, we further compared the short-pulse data with the theoretical predictions. The good agreement between the experimental data and theoretical predictions demonstrated that the experimentally determined parameters of $\Omega_c$, $\alpha$, and $\gamma$ are convincing.

To effectively reduce the contributions of the coupling and pump fields to the single-photon counting modules (SPCMs), we intentionally collected Stokes photons along the coupling propagation direction, and anti-Stokes photons along the pump propagation direction. The reduction scheme will be described in the next two paragraphs, and it attenuated the coupling and pump fields by 147~dB and 157~dB, respectively. Such an arrangement resulted in a phase mismatch, i.e., $L |(\vec{k}_p - \vec{k}_{s} + \vec{k}_c - \vec{k}_{as}) \cdot \hat{z}|$, of about 0.23~rad, where $L$ is the medium length, $\vec{k}_p$, $\vec{k}_c$, $\vec{k}_{s}$, and $\vec{k}_{as}$ are the wave vectors of the pump and coupling fields and the Stokes and anti-Stokes photons, and $\hat{z}$ is the unit vector of the Stokes or anti-Stokes propagation direction. This phase mismatch of 0.23~rad is a very manageable downside, since it makes a negligible reduction of the biphoton's generation rate.

Since the anti-Stokes (Stokes) photons and the pump (coupling) field propagated nearly in the same direction, we installed a polarization filter and a 780-nm bandpass filter (two 795-nm bandpass filters) to prevent the pump (coupling) field from entering the single-mode optical fiber (SMF), collecting the anti-Stokes (Stokes) photons. Each of the polarization filters is the combination of a quarter-wave plate, a half-wave plate, and a polarizing beamsplitter, which reduced the pump (coupling) power by 49~dB (49~dB). The bandpass filter(s) added an additional attenuation of 38~dB (40~dB). After each SMF, we further employed two etalons in series (an etalon) to decrease the pump (coupling) leakage by about 70~dB (58~dB). The total extinction ratio of 157~dB (147~dB) reduced the pump (coupling) power of  56-$\mu$W (1.5-mW) to merely zero (12) photons/s. 

Due to the anti-Stokes (Stokes) photons and the coupling (pump) field having the very similar wavelengths and the same polarization, the above-mentioned polarization and bandpass filters cannot block the leakage of the coupling (pump) field into the anti-Stokes (Stokes) SMF. Fortunately, the angle separation of $1^\circ$ between the anti-Stokes and coupling (between the Stokes and pump) propagation directions had already reduced the leakage significantly. After the SMFs, the two etalons in series (the etalon) further decreased the coupling (pump) leakage by about 60~dB (60~dB). The total extinction ratio of 130~dB (118~dB), including the attenuations due to the angle separation and etalon(s), reduced the coupling (pump) power of 1.5 mW (56 $\mu$W) to about 600 (350) photons/s. Note that the anti-Stokes (Stokes) SPCM (Excelitas SPCM-AQRH-13-FC) has a quantum efficiency of 0.55 (0.56) and a dark count rate of 220 (140)~counts/s.

The two etalons in series for the anti-Stokes photons (the etalon for the Stokes photons) had a net linewidth of 46~MHz (80~MHz), and a peak transmission of around 18\% (30\%). Taking into account of the attenuation due to optics, SMF's coupling ratio, two etalons' total transmission, and SPCM's quantum efficiency, the overall collection efficiency of the anti-Stokes (Stokes) photons was 7.7\% (13\%). Inside the atomic cloud, the Stokes photon propagated at the light speed in vacuum, and the anti-Stokes photon was slow light due to the EIT effect. The Stokes SPCM detected a photon first, and triggered a time tagger (Fast ComTech MCS6A5T8). After some delay time, the anti-Stokes SPCM detected another photon which was recorded as a coincidence count by the time tagger.

\section{Theoretical Model}

The time correlation function between the anti-Stokes and Stokes photons, i.e., the biphoton wave packet, is given by \cite{SFWM.Theory}
\begin{eqnarray}
	G^{(2)}(\tau) &=&
		\left| \int_{-\infty}^{\infty} \frac{d\delta}{2\pi}
		e^{-i\delta\tau}
		\frac{\sqrt{k_{as} k_s}L}{2} \chi(\delta) 
		\right. \nonumber \\ 
	&\times&
		\left.  {\rm sinc} \! \left[ \frac{k_s L}{4} \xi(\delta) \right] 
		e^{i(k_s L/4) \xi(\delta)} \right|^2,
\label{eq:biphoton}
\end{eqnarray} 
where $\tau$ is the delay time of detecting an anti-Stokes photon upon a Stokes photon's trigger, $\delta$ is the two-photon detuning between the Stokes photon and pump field (or $-\delta$ is that between the anti-Stokes photon and coupling field), $k_{as}$ and $k_s$ are the wave vectors of the two photons, $L$ is the medium length, $\chi(\delta)$ is the cross-susceptibility of the anti-Stokes photon induced by the Stokes photon, and $\xi(\delta)$ is the self-susceptibility of the anti-Stokes photon. The formulas of the cross-susceptibility and self-susceptibility are shown below:
\begin{eqnarray}
\label{eq:cross_chi} 
	\frac{\sqrt{k_{as} k_s}L}{2} \chi(\delta) &=&
		\frac{\alpha\Gamma}{4} 
		\frac{\Omega_p}{\Delta_p+ i\Gamma/2} 
		\\
	&\times&	
		\frac{\Omega_c}{\Omega_c^2-4(\delta+i\gamma)(\delta+i\Gamma/2)},~~ 
		\nonumber \\
\label{eq:self_chi}
	\frac{k_s L}{4} \xi(\delta) &=& 
		\frac{\alpha\Gamma}{2} 
		\frac{\delta+i\gamma}{\Omega_c^2-4(\delta+i\gamma)(\delta+i\Gamma/2)},
\end{eqnarray}
where $\alpha$ represents the OD of the entire atoms, $\Omega_p$ and $\Omega_c$ are the Rabi frequencies of the pump and coupling fields, $\Gamma =$ 2$\pi$$\times$6~MHz is the spontaneous decay rate of the excited state, $\Delta_p$ is the detuning of the pump field, and $\gamma$ is the dephasing rate of the ground-state coherence, i.e., the decoherence rate. Since the difference between the spontaneous decay rates of the excited states $|3\rangle$ and $|4\rangle$ is merely about 5\% in our case, we neglect the difference and set the two rates to $\Gamma$.

\section{Results and Discussion}

\FigTwo

We systematically measured the biphoton wave packets or two-photon time correlation functions by varying the coupling powers. The representative data taken at the highest and lowest coupling powers of 1.5~mW and 56~$\mu$W are shown by the circles in Figs.~\ref{fig:biphoton}(a) and \ref{fig:biphoton}(b), respectively. The subfigure (a) [or (b)] is the result of 120,000 (or 105,000) measurements, and each measurement had a time window of 240 $\mu$s. Since we switched off (and on) the MOT during (and after) each measurement, the duty cycle of the biphoton generation is 0.8\%. The width of time bin is 6.4~ns in (a) or 51.2~ns in (b). Considering the overall collection efficiencies of the Stokes and anti-Stokes photons, the generation rates of this biphoton source at the coupling powers of 1.5~mW and 56~$\mu$W were 3460$\pm$140 pairs/s and 3340$\pm$40 pairs/s, respectively. 

The red lines in Fig.~\ref{fig:biphoton} are calculated from the two-photon time correlation function, i.e., $G^{(2)}(\tau)$ in Eq.~(\ref{eq:biphoton}). During the biphoton-generation time window of 240~$\mu$s, the OD of the system gradually decayed to about 90\% of its initial value. We take the OD's decay into account, and all the quoted values of the OD are the average value within this 240~$\mu$s. Since the OD varied merely 10\% and the variation was nearly linear during the measurement, we just use its average value in the evaluation of $G^{(2)}(\tau)$. In the calculation, $\alpha$ and $\gamma$ are set to the values, which are determined by the classical-light data of slow light, and the values of $\Omega_p$ and $\Delta_p$ are also experimentally measured or determined. The value of $\Omega_c$ is adjustable to get the best match of the biphoton's temporal width. Note that $\alpha$, $\Omega_c$, and $\gamma$ affect the profile of the biphoton wave packet, but $\Omega_p$ and $\Delta_p$ do not. In Figs.~\ref{fig:biphoton}(a) and \ref{fig:biphoton}(b), $\Omega_c = 2.1$$\Gamma$ and 0.42$\Gamma$, respectively, while their values determined by the Autler-Townes splitting in the EIT spectrum are $\Omega_c = 2.0$$\Gamma$ and 0.40$\Gamma$. The parameters of $\alpha$, $\gamma$, $\Omega_p$ and $\Delta_p$ (or $\Omega_c$) used in $G^{(2)}(\tau)$ were measured (or verified) by the methods without using the biphoton data. Hence, the red lines are regarded as the theoretical predictions.


The coupling field also drove the far-detuned transition of $|5S_{1/2},F=2,m=1\rangle$ $\rightarrow$ $|5P_{3/2},F=3,m=2\rangle$, and the transition caused the photon switching effect \cite{PS, OurOPL2006}. Consequently, the value of $\gamma$ was approximately linear to that of $\Omega_c^2$ \cite{QM_OurPRL2013, OurPRA88in2013, QM_OurPRL2018}, i.e.,
\begin{equation}
	\gamma = \gamma_0 + a \frac{\Omega_c^2}{\Gamma^2},
\end{equation}
where $\gamma_0$ is the intrinsic decoherence rate of the experimental system and $a$ is the proportionality due to the photon switching effect. The value of $\gamma_0$ was 2$\times$$10^{-4}$$\Gamma$ or 2$\pi$$\times$1.2~kHz. The uncertainty plus day-to-day fluctuation of the decoherence rate is the larger one of $\pm$1$\times$$10^{-4}$$\Gamma$ and $\pm$20\%.

The green lines in Fig.~\ref{fig:biphoton} are the results of four-point moving average, which reduces fluctuation of the data points. Since the temporal full width at the half maximum (FWHM) of 0.57 (or 13.4)~$\mu$s as determined by the red line is much longer than the time-bin width of 6.4 (or 51.2)~ns, the moving average affects the profile of the biphoton wave packet very little. The consistency between the experimental data (the green lines) and the theoretical predictions are satisfactory. We performed the discrete Fourier transform (DFT) on the biphoton wave packets and the Fourier transform (FT) on the theoretical predictions. The baseline or background count of the data was removed before the DFT or FT. The representative spectra are shown in the insets of Fig.~\ref{fig:biphoton}.

The temporal width or spectral linewidth of the biphoton wave packet is mainly determined by the propagation delay time in the experimental system of a large OD and a negligible decoherence rate. This is exactly demonstrated by Fig.~\ref{fig:width}, showing that the FWHM of the biphoton's temporal profile and the corresponding spectral FWHM as functions of $\Omega_c^2$. Under a large OD, the propagation delay time, $\tau_d$ ($= \alpha\Gamma/\Omega_c^2$), is much larger than the inverse of the EIT bandwidth, $\tau_b$ ($= 1/\Delta\omega_{\rm EIT}$ or $\sqrt{\alpha}\Gamma/\Omega_c^2$). Under a negligible decoherence rate, the coherence time, $\tau_c$ ($= 1/\gamma$), is far greater than $\tau_d$. Using the condition of $\tau_c \gg \tau_d \gg \tau_b$, one can show that the biphoton's spectral FWHM, $\Delta \omega/2\pi$, is given by \cite{SFWM.Theory, SFWM.dLambda.ColdAtoms9.Zhang}
\begin{equation}
\label{eq:linewidth}
	\frac{\Delta \omega}{2\pi} \approx \frac{0.88}{\tau_d} = \frac{0.88\Gamma}{\alpha} 
		\left( \frac{\Omega_c}{\Gamma} \right)^2.
\end{equation}
In Fig.~\ref{fig:width}, the blue circles are the experimental data of $\Delta \omega /2\pi$ versus $(\Omega_c/\Gamma)^2$, and the blue line is the best fit of a linear function with zero interception. The initial OD during the biphoton-generation time window, i.e., $\alpha_0$, was 110$\pm$5. Using $\alpha = 110{\pm}5$ in Eq.~(\ref{eq:linewidth}), we obtain $\Delta \omega /2\pi$ = 300$\pm$10~kHz$\times$$(\Omega_c/\Gamma)^2$. The slope of the best fit in Fig.~\ref{fig:width} is 280~kHz. Therefore, the experimental result is very close to the theoretical expectation of a nearly ideal biphoton source with a negligible decoherence rate.  

\FigThree

The longest temporal or narrowest spectral FWHM in Fig.~\ref{fig:width} is 13.4$\pm$0.3~$\mu$s or 50$\pm$1~kHz as demonstrated by Fig.~\ref{fig:biphoton}(b). This is the best record to date and also the first result of the biphoton linewidth below 100~kHz. The very long biphoton had a signal-to-background ratio (SBR) of 3.4, showing that the cross-correlation function between the Stokes and anti-Stokes photons, $g^{(2)}_{s,as}(0)$, is 4.4 \cite{SFWM.dLambda.ColdAtoms10.Du}. Both the auto-correlation function of the Stokes photons, $g^{(2)}_{s,s}(0)$, and that of the anti-Stokes photons, $g^{(2)}_{as,as}(0)$, are approximately equal to 2 \cite{OurPRR2022, AutoCorrelation}. With $g^{(2)}_{s,as}(0) = 4.4$ and  $g^{(2)}_{s,s}(0) = g^{(2)}_{as,as}(0) = 2$, the Cauchy-Schwarz inequality for classical light is violated by 4.8 folds, which clearly deomnstrates the biphoton's non-classicality. Furthermore, by increasing the coupling power we were able to tune the temporal or spectral width to 0.57$\pm$0.02~$\mu$s or 1200$\pm$40~kHz as demonstrated by Fig.~\ref{fig:biphoton}(a). While the temporal or spectral width of the biphoton was shortened or enlarged by about 24 folds, its SBR was significantly enhanced to 40.

\FigFour

We compared the generation rates of the biphoton source operating at different coupling powers or equivalently producing various temporal widths. Under the condition of a large OD and a negligible decoherence rate, i.e., $\Omega_c^2 \gg 2\Delta\omega_{\rm EIT}\Gamma$ and $\Omega_c^2 \gg 2\gamma\Gamma$, the generation rate is approximately proportional to 
\begin{equation}
\label{eq:GR}
	\int_{-\infty}^{\infty} d\tau  G^{(2)}(\tau) \approx 
		\frac{\alpha\Gamma}{2\pi} \frac{\Omega_p^2}{4\Delta_p^2+\Gamma^2} 
		\exp\left( -\frac{\alpha\gamma\Gamma}{\Omega_c^2} \right).
\end{equation}
As long as the value of $\alpha\gamma\Gamma/\Omega_c^2$ is small, varying the coupling power or tuning the biphoton temporal width changes the generation rate a little. This is exactly the case in our experiment. Although we changed the biphoton temporal width by about 24 folds, the generation rate varied in the range of 3300$\sim$3800 pairs/s or fluctuated merely about $\pm$7\% as shown by Fig.~\ref{fig:brightness}. That is, we generated either a longer biphoton pulse with a smaller amplitude or a shorter biphoton pulse with a larger amplitude, while the area below the pulse or the pulse energy was approximately the same \cite{OurPRA72in2005}. Only a negligible decoherence rate in the EIT system can make it happen.

The brightness is defined as the generation rate per pump power, and the spectral brightness is defined as the brightness per linewidth. Figure~\ref{fig:brightness} shows the brightness and the spectral brightness as functions of $\Omega_c^2$. During the measurement, the pump power was kept the same, and thus the behavior of the brightness is essentially the same as that of the generation rate. We considered the fluctuations of the OD and the decoherence rate, and utilized the predictions of Eq.~(\ref{eq:GR}) to plot the gray area. The consistency between the data and the predictions is satisfactory. Furthermore, since the brightness varied a little against $\Omega_c^2$ and the biphoton spectral width is linearly proportional to $\Omega_c^2$, the spectral brightness or generation rate per pump power per spectral linewidth approximately depends on $1/\Omega_c^2$. Such a dependence is clearly seen in Fig.~\ref{fig:brightness}. In this work, the highest spectral brightness is 1.2$\times$$10^6$ pairs/(s$\cdot$mW$\cdot$MHz).

\section{Conclusion}

In conclusion, we were able to produce the narrow-linewidth biphoton with a temporal FWHM of 13.4~$\mu$s, corresponding to a spectral FWHM of 50~kHz, owing to a large optical depth and a negligible decoherence rate in the experimental system. As for a wide range of the Rabi frequency or power of the coupling field employed in this work, the large optical depth maintained the criterion that the propagation delay time is far greater than the reciprocal of the EIT bandwidth, and the negligible decoherence rate preserved the criterion that the coherence time is much longer than the propagation delay time. Consequently, not only the temporal profile of the biphoton wave packet was predominately determined by the propagation delay time, but also we tuned the temporal or spectral width by about 24 folds and, at the same time, kept the generation rate approximately the same. This work has demonstrated a high-efficiency ultranarrow-linewidth biphoton source, and achieved the milestone of an ultralong biphoton wave packet. 

\section*{Acknowledgments}
This work was supported by the Ministry of Science and Technology of Taiwan, under Grants No. 109-2639-M-007-002-ASP and No. 110-2639-M-007-001-ASP.



\end{document}